\documentclass[conference]{IEEEtran}
\IEEEoverridecommandlockouts

\usepackage{cite}
\usepackage{amsmath,amssymb,amsfonts}
\usepackage{algorithmic}
\usepackage{graphicx}
\usepackage{subfigure}
\usepackage{textcomp}
\usepackage{url}
\usepackage{caption}
\usepackage{tabularx}
\usepackage{makecell}
\usepackage{diagbox}
\usepackage{algorithm}
\usepackage{xcolor}
\def\BibTeX{{\rm B\kern-.05em{\sc i\kern-.025em b}\kern-.08em
    T\kern-.1667em\lower.7ex\hbox{E}\kern-.125emX}}
\begin{document}

\title{Practical Spoofing Attacks against Galileo OSNMA with Time-Synchronization Manipulation}
\author{Haiyang Wang, 
	Yuanyu Zhang,
	Yangke Tan, 
	Ji He,
	Shuangrui Zhao,
	Ning Xi,
	and Yulong Shen
	\thanks{This work was supported in part by the National Natural Science Foundation of China (No. 62220106004, 92467201), in part by the Fundamental Research Funds for the Central Universities (Grant No. QTZX25080, ZDRC2202), in part by Shaanxi Elite Talent Introduction Program (Youth Project), in part by the Young Talent Fund of Association for Science and Technology in Shaanxi, China. 
    (\emph{Corresponding author}: Yuanyu Zhang)}
	\thanks{H. Wang, Y. Zhang, Y. Tan, J. He, S. Zhao, N. Xi and Y. Shen are with the School of Computer Science \& Technology, Xidian University, Xi'an, Shaanxi 710071, China. Email: \{hywang\_xdu, yangketan\}@stu.xidian.edu.cn, \{yyuzhang, jihe, zhaoshuangrui, nxi\}@xidian.edu.cn, ylshen@mail.xidian.edu.cn.}
}

\maketitle

\begin{abstract}
Galileo launched the Open Service Navigation Message Authentication (OSNMA) to defend against spoofing attacks. 
This paper identifies an artificially manipulated time synchronization (ATS) condition in OSNMA-enabled receivers, under which attackers can jointly manipulate the Galileo signals and a receiver’s local reference time (LRT) while still satisfying the time synchronization (TS) requirement. 
Exploiting the ATS condition, we propose a TS-compliant spoofing framework, including TS-compliant replay (TSR), TS-compliant forgery (TSF), and TS-compliant dual-frequency forgery (TSDF) attacks. 
The principle underlying these attacks is to first force the receiver to satisfy the TS requirement by manipulating its LRT, and then transmit carefully designed signals to spoof the receiver to the attacker-selected locations and times.
More specifically, the TSR attack replays previously recorded E1 signals, the TSF attack transmits forged E1 signals containing valid OSNMA data, and the TSDF attack simultaneously forges both E1 and E5b signals, targeting dual-frequency receivers that support cross-band authentication.
To validate the effectiveness of the proposed attacks, we conduct real-world experiments using two commercial Galileo receivers and two open-source software-defined radio (SDR) receivers that support OSNMA. 
The results showed that all attacks can successfully pass OSNMA authentication and spoof receivers to attacker-selected locations and times.
\end{abstract}


\section{Introduction}
Galileo is a major Global Navigation Satellite System (GNSS), providing global positioning, navigation, and timing services.
Because the structures of civilian signals are public, attackers can forge signals to mislead receivers into calculating false positions and times \cite{uavtakover-usenix54,noah2025gnss-survey}. 
Fortunately, the Galileo system officially launched Open Service Navigation Message Authentication (OSNMA), providing a certain level of protection \cite{simon2025galileo,galan2025TTFAF}.
OSNMA is currently the only operational GNSS navigation message authentication service. 
Leading GNSS receiver manufacturers such as Septentrio and u-blox have integrated OSNMA support into their products covering a variety of applications, including automotive systems, industrial automation, UAVs, and robotics \cite{OSNMA-devices,OSNMA-Ublox1}.


At the core of OSNMA is the Timed Efficient Stream Loss-Tolerant Authentication (TESLA) protocol, which computes a message authentication code (MAC) for each navigation data using a key obtained from a global TESLA key chain \cite{gotzelmann2023OSNMA58}. Each key in the chain is associated with a Galileo System Time (GST).
The navigation data and MAC are broadcast immediately to receivers, while the key is disclosed later. 
Only after receiving the corresponding key can a receiver authenticate the navigation data.
To ensure OSNMA security, receivers must comply with a mandatory time synchronization (TS) requirement, i.e., the difference between a receiver's local reference time (LRT) and the GST extracted from received Galileo signals must not exceed a given threshold \cite{OSNMAreceiver21}. 
A key with GST violating the TS rule is viewed as outdated and will be discarded.
The above mechanisms of delayed key disclosure, navigation data verification, and TS check together empower Galileo with the ability to resist spoofing attacks \cite{o2025galileo}. 

However, OSNMA does not eliminate all security threats against Galileo signals.
Several studies have explored attacks against OSNMA-protected signals, including replay-based attacks such as distance-decreasing (DD), and pre-startup replay (PreRep) attacks \cite{DDAttack31, wang2023novel52}, proximity-based replay attacks \cite{lazyreplay2023, distributereplay40}. 
Although the above works demonstrate the possibility of attacking OSNMA, they remain limited in several aspects. 
First, replay-based attacks require signal recording and replaying between colluding attackers, and can only spoof the location to the original recorded location of the signal.
Second, these attacks typically require real-time forwarding of signal samples with low latency to satisfy the TS constraint.
Another limitation is that existing attacks mainly focus on the Galileo E1 signal. However, OSNMA data transmitted on E1 can also authenticate navigation data received on E5b through cross-band authentication. Therefore, dual-frequency receivers may easily detect such attacks.
More importantly, most existing attacks are evaluated using receivers that do not support OSNMA, so the feasibility of these attacks against OSNMA-integrated receivers remains uncertain.

This paper conducts an in-depth investigation of the OSNMA mechanism and the TS implementation on the receiver side, and identifies an artificially manipulated time synchronization (ATS) condition.
Under this condition, attackers can jointly manipulate the receiver's LRT and the GST in the spoofing signal, forcing the receiver to pass the TS check. 
Based on the ATS condition, we propose three practical spoofing attacks against OSNMA-enabled receivers, including TS-compliant replay (TSR), TS-compliant forgery (TSF), and TS-compliant dual-frequency forgery (TSDF) attacks.
The basic idea is to first tamper with the receiver's LRT to meet the TS rule, and then transmit carefully designed spoofing signals. 
TSR attack eliminates the limitations of real-time relay by aligning the receiver's LRT with the GST of a previously recorded Galileo signal. 
TSF attack allows attackers to select target locations and times to forge and transmit E1 signals containing valid OSNMA data. 
TSDF attack further forges consistent E1 and E5b signals to spoof dual-frequency receivers that support cross-band authentication. 
We implemented all three attacks using software-defined radio (SDR) hardware and evaluated them against two commercial receivers and two SDR-based receivers, all of which support OSNMA. 
We further discuss enhancement strategies for improving OSNMA security.
The main contributions of this paper are summarized as follows:
\begin{itemize}
\item[$\bullet$]  \textbf{ATS condition}. We identify and experimentally demonstrate an exploitable ATS condition in OSNMA-enabled receivers, under which attackers can manipulate the GST in the Galileo signals and LRT of a victim receiver to meet the TS rule.
The condition lays the foundation for the attacks proposed in this paper.

\item[$\bullet$] \textbf{TS-compliant spoofing framework}. Building upon the ATS condition, we propose TSR, TSF, and TSDF attacks. 
These attacks all first tamper with the receiver's LRT to meet TS requirements.
Then, TSR replays previously recorded E1 signals to spoof the receiver to the location and time of the signal recording.
TSF forges E1 signals embedded with valid OSNMA data, which can spoof the receiver to attacker-selected locations and times. 
Extending TSF to dual-frequency receivers, TSDF forges consistent E1 and E5b signals to satisfy OSNMA cross-band authentication.
\item[$\bullet$] \textbf{Real-world attack implementation.}
We conducted extensive real-world experiments and evaluations of the proposed attacks using OSNMA-enabled receivers, including two commercial receivers and two SDR-based receivers.
The experimental results showed that all attacks can successfully pass OSNMA authentication and spoof receivers to attacker-selected locations and times.
\end{itemize}

The remainder of the paper is organized as follows. In Section \ref{sec:related-work}, we review the related work on GNSS spoofing attacks. In Section \ref{sec:OSNMA-Analysis}, we analyze the OSNMA mechanism in detail and present the ATS condition. Section \ref{sec:attack} introduces the proposed TSR, TSF and TSDF attacks. Section \ref{sec:Impl} demonstrates the implementation of these attacks in real-world environments and analyzes the experimental results. 
Section \ref{sec:discuss} discusses OSNMA enhancement strategies.
Finally, Section \ref{sec:conclusion} concludes the paper.

\section{Related Work}\label{sec:related-work}
We divide the work related to spoofing attacks on OSNMA into two categories: forgery attacks and replay attacks, with a primary focus on work conducted in real-world environments.
\subsubsection{Forgery Attacks}
Forgery attacks on OSNMA are currently unexplored, while those on unauthenticated GNSS signals without considering OSNMA are widely studied.  
For example, researchers at the Mobile Security of Alibaba Group successfully spoofed the time and location of smartphones and smartwatches using open-source software and SDR \cite{wanggpsspoof25}. 
Sathaye \emph{et al.} developed a real-time GPS signal generator to spoof and take over commercial UAVs \cite{uavtakover-usenix54}.
Tibaldo \emph{et al.} proposed a wide-area GNSS spoofing technique that preserves the relative distances among receivers within the spoofed region \cite{tibaldo2025gnss}.
Humphreys \emph{et al.}  proposed a more covert attack scheme, placing a receiver-spoofer near the victim receiver to estimate the receiver’s position and velocity \cite{forgeryAttack41}.  
Based on these estimates, the spoofer aligns its counterfeit signals with the authentic signals and gradually increases their power until the receiver locks onto them.

Obviously, the above attacks fail to work on OSNMA-integrated Galileo receivers because they did not take the OSNMA scheme into consideration and will be easily detected.

\subsubsection{Replay Attacks}
Many researchers have conducted experiments on replay attacks in real-world environments. 
For example, researchers from the KTH Royal Institute of Technology conducted experiments based on two colluding adversaries \cite{replay12}: one recorded authentic GNSS signals, and the other replayed these signals at a different location via networks.
Zhang \emph{et al.} proposed a DD attack that estimates the early portion of each Galileo bit or symbol and begins transmitting before reception is complete \cite{DDAttack31}. 
This attack reduces the victim’s computed pseudorange, but symbol errors can corrupt authenticated navigation data and trigger OSNMA verification failures.

The proximity-based replay attack in \cite{lazyreplay2023} assumes that the attacker and victim observe the same Galileo navigation messages, allowing the attacker to manipulate the code phase and Doppler shift for spoofing. 
The attack was tested on a commercial receiver (ublox M8N) that does not support OSNMA.
Wang \emph{et al.} proposed a real-time PreRep attack, which controls the replay delay to within 30 seconds to satisfy the TS requirement, and demonstrated it against an OSNMA-enabled receiver \cite{wang2023novel52}.

\section{OSNMA and TS Analysis}\label{sec:OSNMA-Analysis}
This section analyzes the OSNMA mechanism and introduces the ATS condition.
\subsection{OSNMA Basics}\label{sec:OSNMA Basics}
\subsubsection{Galileo Subframe}\label{sec:message-format}
\begin{figure}[h]
\centerline{\includegraphics[width=\linewidth]{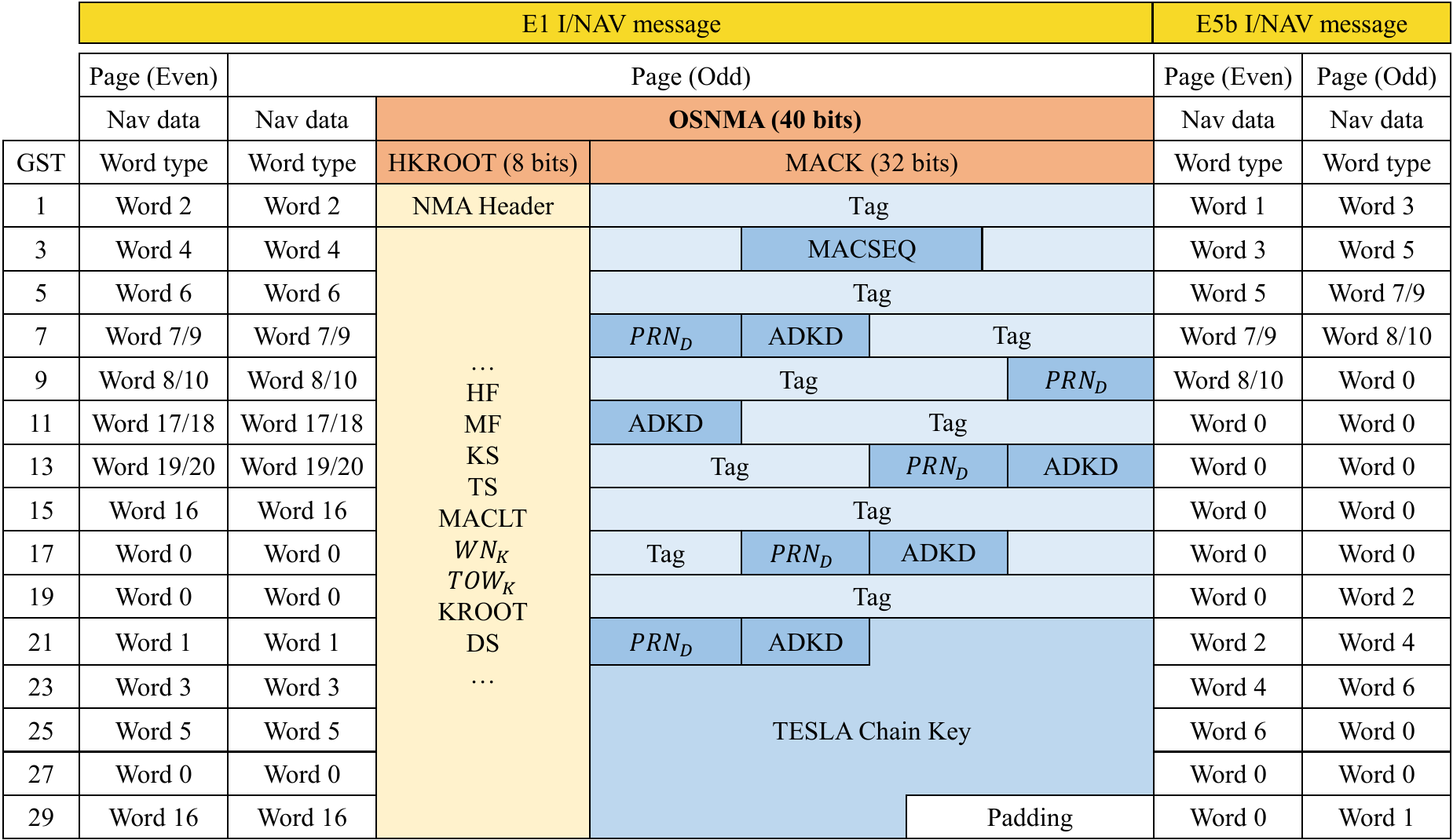}}
\caption{Structure of E1/E5b subframe and OSNMA fields.}
\label{fig:subframe}
\end{figure}

The Galileo system broadcasts I/NAV messages simultaneously in the E1 and E5b bands, organized into subframes \cite{galileosis}. 
Each subframe lasts 30 seconds and consists of 15 pages. 
OSNMA data and navigation data are contained within the corresponding fields of each page, as shown in Fig. \ref{fig:subframe}.
Navigation data are categorized into different word types, each containing specific parameters, such as ephemeris data and Galileo System Time (GST), used for positioning and timing.
The 40-bit OSNMA field is broadcast within the E1 signal and is divided into an HKROOT section and a MACK section. 
A receiver combines 15 MACK sections within a subframe to obtain a 480-bit MACK message, and extract a TESLA key and tags (i.e., MACs). 
The receiver similarly combines HKROOT sections across subframes to recover the digitally signed material used to authenticate the TESLA root key.

\subsubsection{TESLA Key Verification}
All Galileo satellites maintain the same TESLA chain, which stores all the keys used to generate the tags.
This chain starts with a random seed key $K_{N}$, known only to the OSNMA provider, and ends with a root key $K_{0}$. 
Each key (say $K_i$, $i\in{0,1,2,\cdots, N-1}$) is computed from its previous key (i.e., $K_{i+1}$) through a hash function, i.e., $K_i=Hash(K_{i+1})$.
Keys are disclosed in the reverse order of their generation, preventing an attacker from deriving an undisclosed key from a disclosed one. 
After verifying the signed root key with Galileo’s public key, a receiver authenticates a disclosed $K_i$ by repeatedly hashing it toward $K_{0}$. 
The required number of hash operations follows from the GST indices of the two keys.

\subsubsection{Tag Verification}\label{sec:tag-verification}
OSNMA assigns an Authentication Data and Key Delay (ADKD) type to each tag to specify which navigation data are authenticated and which disclosed key should be used. 
For a tag with ADKD=0 in the $i+1$ subframe, the receiver authenticates words 1–5 from the $i$ subframe using $K_{i+2}$.
ADKD=4 combines words 6 and 10 and also uses $K_{i+2}$.
ADKD=12 combines the same data as ADKD=0 but uses $K_{i+12}$.
All ADKD types use the same tag generation algorithm:
\begin{equation}\label{eq1}
	tag = trunc (TS,MF(K, m_a)),
\end{equation}
where $\operatorname{trunc}$ truncates the output to tag size (e.g., $TS$), $MF$ denotes a MAC function, and the $K$ depends on the ADKD type. 
The authenticated message $m_a$ is computed as:
\begin{equation}\label{eq2}
	m_a = ({PRN}_{D}||{PRN}_{A}||{GST}_{sf}||CTR||{navdata}||P),
\end{equation}
where ${PRN}_{D}$ and ${PRN}_{A}$ represent the satellite transmitting navigation data and the satellite transmitting OSNMA data, respectively. 
$CTR$ is the tag's position in the MACK message, and $navdata$ denotes the navigation data specified by the ADKD type.
If the calculated tag matches the received tag, the receiver marks the navigation data as authentic.

Because some Galileo satellites do not broadcast OSNMA data, Galileo system employs a cross-satellite authentication mechanism to authenticate their navigation data \cite{OSNMAsisicd20,galan2025sensitivity}.
Specifically, a satellite ${PRN}_{A}$ that transmits OSNMA broadcasts tags computed based on the navigation data from another satellite ${PRN}_{D}$.
The MAC lookup table (MACLT) in HKROOT declares a tag sequence that determines each tag’s position in the MACK message, ADKD type, and authentication scope (self or cross-satellite). 
For example, when MACLT=34, the sequence is “00S, FLX, 04S, FLX, 12S, 00E”, which defines six tag slots in the MACK message. “00S”, “04S”, and “12S” are self-authentication tags with the indicated ADKD values. “00E” is a cross-satellite tag with ADKD=0 and FLX denotes flexible allocation.

In addition to cross-satellite authentication, OSNMA also supports cross-band authentication. 
Galileo E1-B and E5b-I carry consistent I/NAV data at the subframe level, as shown in Fig. \ref{fig:subframe}.
A dual-frequency receiver can use OSNMA data from E1 band to authenticate navigation data received on both E1 and E5b.

\subsection{ATS Condition}\label{sec:ATS condition}
OSNMA requires receivers to meet the mandatory TS requirement \cite{OSNMAreceiver21}, i.e., the difference between the true time (denoted as $t_{true}$) and the GST (denoted as $t_{GST}$) must not exceed a threshold $T_L$, which is typically 30~s.
Only when the TS requirement is satisfied can the OSNMA mechanism initialize and authenticate navigation data.
Because true time is not directly observable, the receiver instead uses an LRT (denoted as $t_{LRT}$), which can be obtained from an internal real-time clock (RTC), a host, or a network time server.
The LRT may differ from true time because of accumulated clock drift or network delay. 
OSNMA models this error as bounded:
\begin{equation}\label{eq4}
	t_{true} \in \left[ {t_{LRT} - B,t_{LRT} + B} \right],
\end{equation}
where $B$ denotes the error bound, the value of which is determined by the receiver manufacturer and configuration.
Introducing parameter $B$ allows the receiver to adopt a practical TS check \cite{OSNMAreceiver21}, defined as: 
\begin{equation}\label{eq5}
	\left|t_{\mathrm{LRT}}-t_{\mathrm{GST}}\right|<B,
\end{equation}

The TS rule in Eq.~\eqref{eq5} provides a certain level of resistance against replay and forgery attacks. 
The GST of a replayed signal typically lags significantly behind the LRT value, whereas a forged signal relies on previously disclosed TESLA keys, the corresponding GST of which are already outdated.
Consequently, such signals are rejected during the TS procedure.

However, the effectiveness of the TS rule fundamentally relies on the correctness of the LRT. 
For receivers that obtain the LRT from an RTC, the accumulated clock drift may become significant after long offline periods. 
Environmental factors, oscillator instability, and imperfect calibration can further enlarge the deviation between the LRT and the true time \cite{2020independent21}.
For those obtaining the LRT from external time servers, the situation can be even worse. 
If the synchronization protocol lacks strong authentication, an attacker may manipulate time responses. 
Examples include unauthenticated Network Time Protocol (NTP) \cite{ntpAttack46}, simple NTP \cite{sntp-attack}, and Precision Time Protocol (PTP) deployments \cite{PTP-attack}.
Some receivers also use host or manufacturer-specific time commands to set the LRT.
If access to these interfaces is not adequately protected, an attacker can also tamper with the LRT \cite{OSNMA-Ublox1}.

\begin{figure}[h]
\centerline{\includegraphics[width=\linewidth]{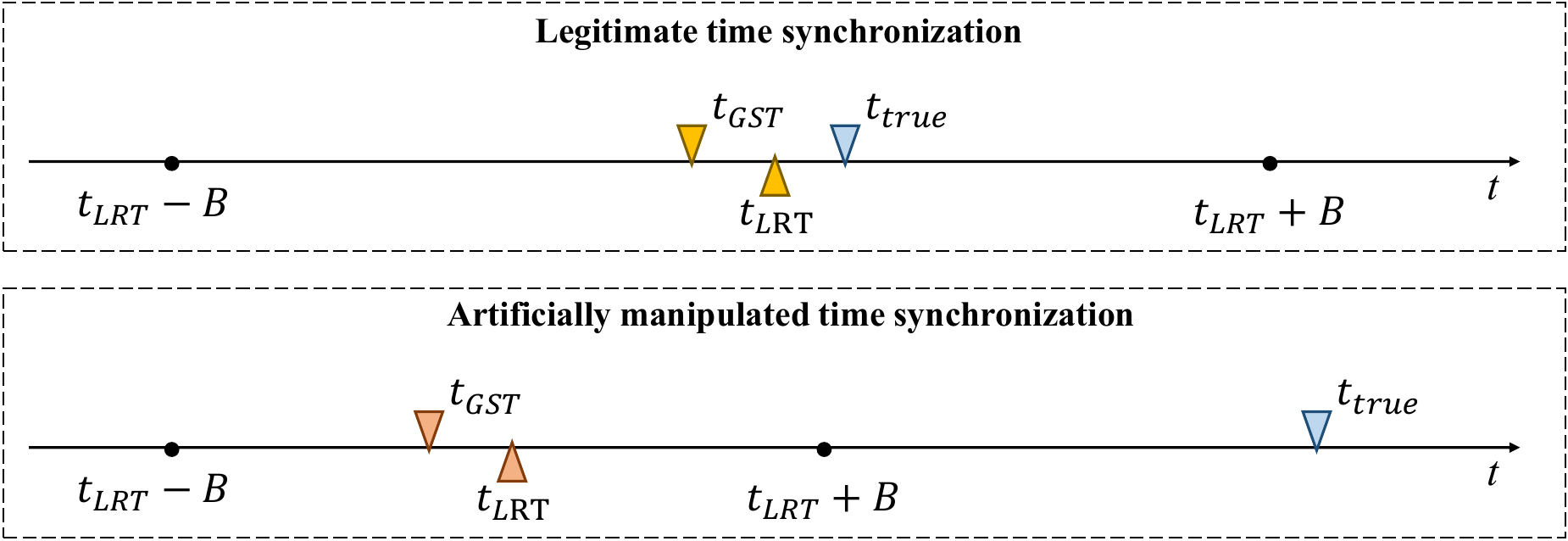}}
\caption{ATS condition illustration. The manipulated GST and LRT meet the TS requirement, but both are far behind the true time.}
\label{fig:ATS}
\end{figure}

Therefore, the TS procedure verifies only the consistency between GST and LRT, rather than the independent correctness.
If a receiver obtains its LRT from an insecure source, an attacker can jointly manipulate the LRT and the GST in the spoofing signals so that they remain mutually consistent and satisfy the TS requirement, even though both deviate from true time, as shown in Fig.~\ref{fig:ATS}.
We refer to this receiver-side condition as Artificially Manipulated Time Synchronization (ATS).
ATS does not compromise OSNMA's cryptographic primitives.
Instead, it exploits the receiver's reliance on a potentially manipulable LRT to cause the TS check to accept mutually consistent but incorrect time values.
ATS therefore serves as the basis for the spoofing attacks presented in this paper.

\section{Proposed Attacks} \label{sec:attack} 
Based on ATS condition, this section introduces a TS-compliant spoofing attack framework against OSNMA, including TSR, TSF, and TSDF attacks.

\subsection{Threat Model}\label{sec:ThreatModel} 
The attacker's goal is to make OSNMA-enabled receivers track the spoofing signal and calculate the attack-selected position, velocity and time (PVT) solutions, which is successfully authenticated by OSNMA. 
We assume the attacker can transmit spoofed Galileo signal near the victim and has sufficient transmission power to suppress the authentic Galileo signal, as in conventional GNSS spoofing.
The attacker can collect broadcast Galileo signals and extract OSNMA data, but cannot deduce the undisclosed key, forge a signed root key, or compromise the OSNMA cryptographic primitives.

Before or during TS procedure, the attacker can tamper with the receiver's LRT, depending on how the LRT is acquired.
As discussed in Section \ref{sec:ATS condition}, possible attack paths include unauthenticated network time protocols and insufficiently protected host or receiver time-setting commands.
For example, as one of the most widely used synchronization protocols, NTP is vulnerable to man-in-the-middle (MitM) attacks \cite{Ntp-security45,NtpAnalysis44,ntpattack47}. 
By forging NTP response packets, an attacker can modify the LRT obtained from a public time server.
Signal spoofing and LRT manipulation are independent, we combine them to spoof OSNMA-protected receivers.

\subsection{TS-compliant Replay (TSR) Attack}\label{sec:TSR attack} 
The key idea of TSR is to replay previously recorded Galileo signals while simultaneously manipulating the victim receiver's LRT to satisfy the TS rule. 
Unlike existing replay attacks that rely on real-time forwarding of signal samples between colluding attackers, TSR can be implemented by a single attacker using non-real-time signal replay. 
As a result, the attack eliminates the strict bandwidth and latency requirements of traditional replay attacks.

Without LRT manipulation, the GST contained in recorded signals is typically far behind the receiver's LRT and therefore violates the TS rule. 
Consequently, although the replayed signals contain authentic OSNMA data, the receiver will reject them during TS. 
By exploiting ATS, the attacker artificially modifies the receiver's LRT to match the GST of the replayed signals, causing the TS procedure to succeed.
The TSR attack is performed as follows:
\begin{enumerate}
\item The attacker records authentic Galileo signals at an arbitrary time.
\item Near the victim, the attacker estimates the time offset between the LRT and the GST in the recorded signal.
\item The attacker aligns the victim receiver's LRT with the recorded GST and replays the recorded signals. 
\end{enumerate}
Since the GST now satisfies Eq. \ref{eq5}, the receiver processes the replayed signal for positioning and OSNMA authentication.

\subsection{TS-compliant Forgery (TSF) Attack}\label{sec:TSF attack}
TSR attack can only spoof the receiver to the location and time where the signal was recorded.
TSF removes the location constraint by forging Galileo E1 signals for an attacker selected target position and time and embedding forged OSNMA data that authenticate the forged navigation data. 
By exploiting ATS, the attacker then aligns the victim’s LRT with the forged GST. 
In TSF attack, the target location can be arbitrarily specified, but the target time can only be set to the past, i.e., the time when the OSNMA authentication materials were disclosed.

For $M\geq4$ tracked satellites, a receiver estimates its position $\mathbf{x}_r$ and clock bias $b_r$ using the algorithm \cite{understandGPS61}: 
\begin{equation}
\label{eq:positioning}
\rho_j = \lVert\mathbf{x}_r-\mathbf{x}_j\rVert_2+c b_r, \quad j=1,\ldots,M,
\end{equation}
where $\mathbf{x}_j$ is the position of $j-th$ satellite derived from its navigation data, $\rho_j$ is the pseudorange derived from the pseudorandom-noise (PRN) code phase, and $c$ is the speed of light.

Position spoofing can be viewed as the inverse problem of GNSS positioning.
Given a target position and time ($x_{r}$, $y_{r}$, $z_{r}$, $t_{r}$),
the attacker synthesizes more than four Galileo satellite signals that produce the desired position and time when processed by the receiver.
Specifically, the attacker determines the visible satellites at ($x_{r}$, $y_{r}$, $z_{r}$, $t_{r}$), computes the corresponding satellite coordinates and pseudoranges, and derives the corresponding code phases, navigation messages, and Doppler shifts. 
These parameters are then used to synthesize Galileo E1 signals are consistent with the desired navigation solution.
The forged navigation messages carry forged GST parameters consistent with $t_{r}$ to spoof the receiver's time solution.
Naturally, such forged signals do not contain valid OSNMA data.

Generating forged Galileo signals is only the first step. 
The main challenge of TSF is reconstructing complete OSNMA authentication relationships for forged navigation messages, including root key, TESLA key authentication, tag self-authentication, and cross-satellite authentication.
The attacker cannot create a new signed root key or derive undisclosed TESLA keys. 
Instead, TSF attack reuses previously broadcast OSNMA material, which we call auxiliary OSNMA data.
These data can be extracted from previously broadcast Galileo navigation messages or from the ESA public repository \cite{OSNMAreceiver21}. 
Because all Galileo satellites share a TESLA chain, auxiliary OSNMA data collected from any ${PRN}_{A}$ satellite can provide the root key and disclosed TESLA keys. 
Provided that the GST associated with the reused keys matches the forged GST, the receiver can successfully authenticate the root key and TESLA key chain. 
Furthermore, the TESLA keys enable the generation of valid tags for forged navigation messages.

Another challenge arises from OSNMA's cross-satellite authentication mechanism.  
Since only a subset of Galileo satellites transmit OSNMA data (see Section \ref{sec:tag-verification}), when forging OSNMA data for ${PRN}_{A}$, an attacker must generate both the self-authentication tags and cross-satellite tags for non-OSNMA satellites.
This introduces an allocation problem: which satellites should be authenticated by ${PRN}_{A}$, and how should the corresponding tags be assigned to the available MACK tag slots? 
Because the OSNMA allocation policy is not public, the attacker must construct a valid custom allocation.

In this paper, we adopt a distance-based tag allocation strategy.
The attacker selects the satellites that will not transmit OSNMA data, and the number of these satellites should not exceed the number of cross-satellite tag slots in the MACK message.
For each non-OSNMA satellite, the inter-satellite distance to the transmitting satellite ${PRN}_{A}$ is computed, and the corresponding cross-satellite tags are assigned in ascending order of distance. 
Satellites that are geographically closer are more likely to be simultaneously visible to the receiver.

After obtaining auxiliary OSNMA data and determining the allocation strategy, the attacker reconstructs a complete OSNMA authentication structure for each forged satellite and embeds it into the forged navigation messages.
\subsubsection{Authentication Parameter Extraction}\label{sec:ParameterExtraction} 
The attacker first reconstructs the authentication environment associated with the forged GST. 
Specifically, time-consistent TESLA keys are extracted from auxiliary OSNMA data to build a local key chain, while HKROOT messages are parsed to extract the MACLT, ADKD type, MAC function, and tag size required for subsequent tag generation.
\subsubsection{Self-Authentication Reconstruction}\label{sec:self-authen}
Using the recovered parameters and local key chain, the attacker generates the self-authentication tags  for each forged ${PRN}_{A}$ satellite using Eq. \ref{eq1}.
\subsubsection{Cross-Satellite Authentication Reconstruction}\label{sec:cross-authen}
The attacker assigns non-OSNMA satellites to ${PRN}_{A}$ satellites according to the inter-satellite distance, and then generates the corresponding cross-satellite tags.
\subsubsection{OSNMA Field Construction}\label{sec:ConstructeOSNMA}
The attacker initializes empty OSNMA fields and populates them with reused HKROOT sections, TESLA keys, self-authentication tags, associated PRNs according to the recovered tag sequence.
Then, based on inter-satellite distance results, the attacker embeds the cross-satellite tags into the corresponding slots in ascending order of distance. 
\subsubsection{Navigation Message Assembly}
\begin{figure}[h]
\centerline{\includegraphics[width=\linewidth]{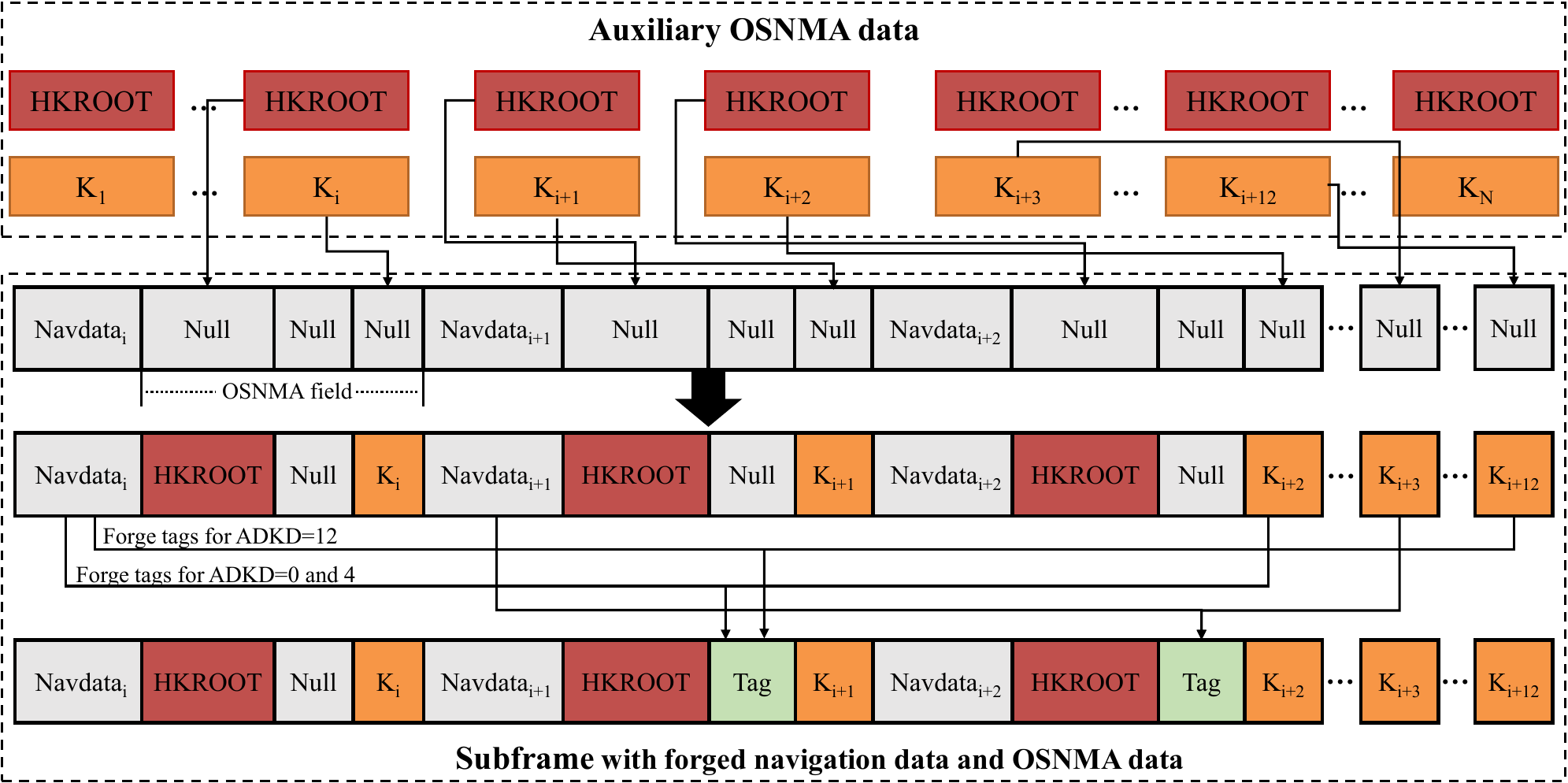}}
\caption{Construction of the forged OSNMA data field.}
\label{fig:OSNMA-constuct}
\end{figure}

Finally, the constructed OSNMA data are embedded into forged I/NAV pages together with the forged navigation data, as shown in Fig. \ref{fig:OSNMA-constuct}. 
The attacker recomputes cyclic redundancy check (CRC) values for each page and assembles complete subframes for transmission.

The resulting navigation messages satisfy valid OSNMA authentication relationships and can successfully pass root key, TESLA key authentication, and tag authentication. 
Combined with ATS-based LRT manipulation, the forged E1 signal can spoof the victim receiver to an arbitrary location and to a past time for which the required authentication material has been disclosed.

\subsection{TS-compliant Dual-Frequency Forgery (TSDF) Attack}\label{sec:TSDF attack}
The TSF attack enables arbitrary position spoofing against receivers that rely solely on Galileo E1 signals. 
However, modern Galileo receivers increasingly employ dual-frequency positioning using both E1 and E5b signals. 
In such receivers, OSNMA data transmitted on E1 can authenticate navigation data received on E5b through cross-band authentication. 
Consequently, if an attacker forges only E1, authentic E5b observations will be inconsistent with the forged E1 measurements and navigation data, making the attack easily detectable.

\begin{figure}[h]
\centerline{\includegraphics[width=\linewidth]{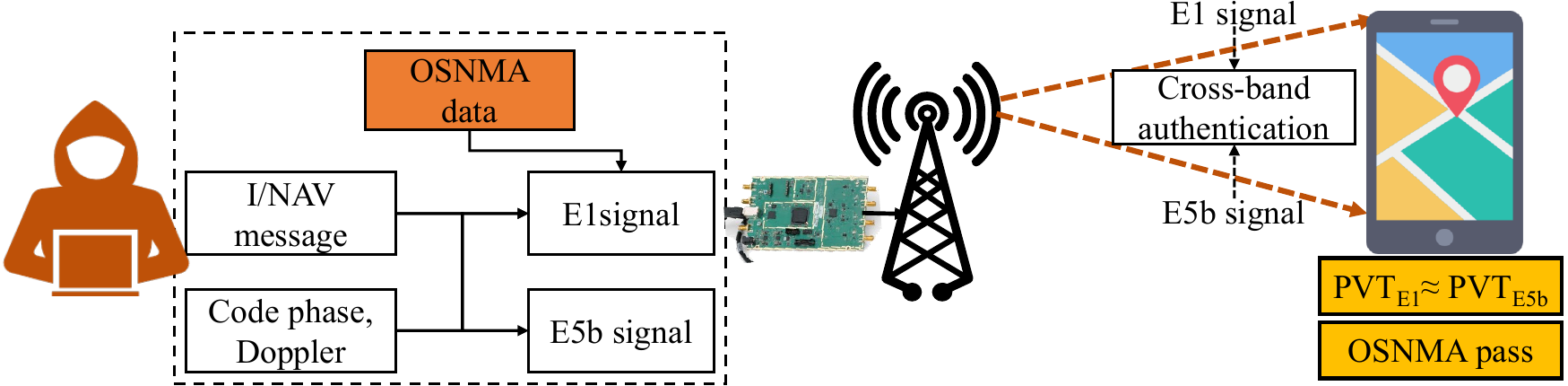}}
\caption{TSDF attack against E1 and E5b cross-band authentication.}
\label{fig:TSDF-Attack}
\end{figure}
To address this limitation, we propose a TSDF attack, which simultaneously forges E1 and E5b signals while preserving consistency across both authentication and positioning.
First, the attacker must ensure navigation message consistency between E1 and E5b. 
Specifically, the forged E1 and E5b navigation data must contain mutually consistent ephemeris and GST so that E1 tags authenticate the E5b navigation data.
The attacker must also ensure measurement consistency across frequencies. 
Specifically, the forged E1 and E5b signals must produce pseudorange observations that correspond to the same target position and time. 
This requires consistent satellite geometry, code phases, and Doppler shifts across both frequency bands, such that the receiver derives matching PVT solutions from E1 and E5b, as shown in Fig. \ref{fig:TSDF-Attack}.

The attacker then simultaneously forges and transmits E1 signals containing OSNMA data and E5b signals.
With LRT manipulation, both frequency signals successfully satisfy the TS rule and cross-band authentication, causing the receiver to derive a forged dual-frequency navigation solution.

\section{Attack Implementation and Evaluation} \label{sec:Impl}
\begin{figure}[h]
    \centering
    \subfigure[]{
        \includegraphics[width=0.22\linewidth]{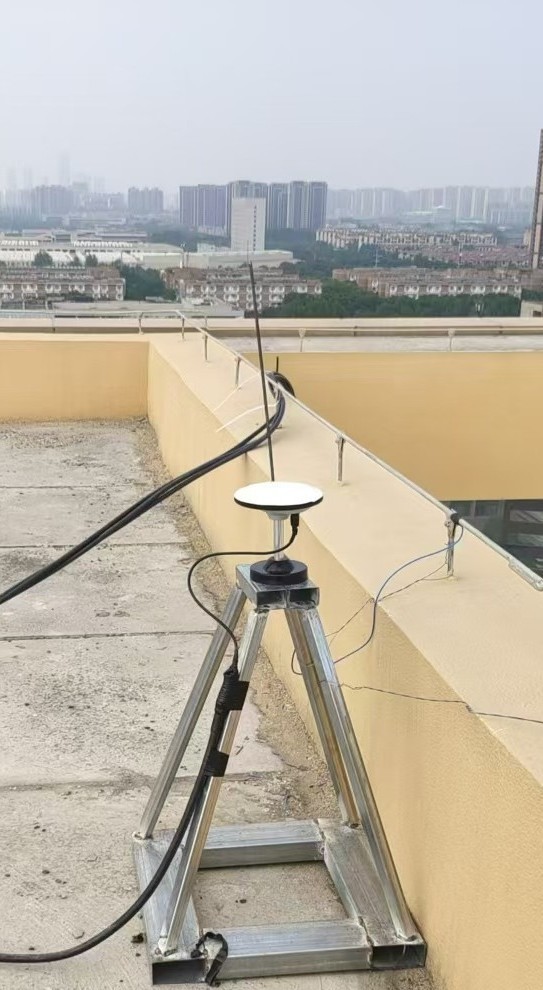}
        \label{fig:antenna}
    }
    \subfigure[]{
        \includegraphics[width=0.69\linewidth]{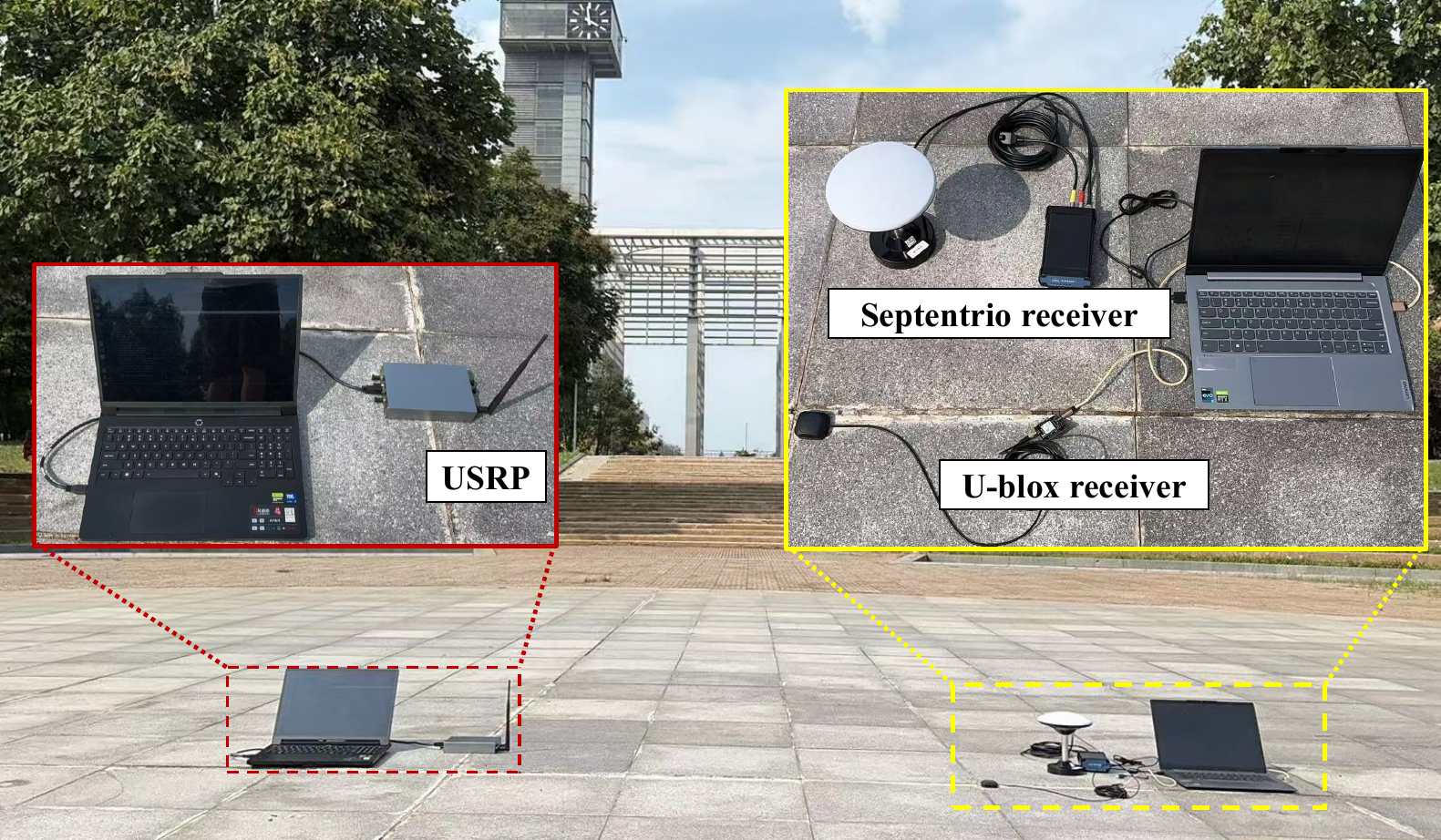}
        \label{fig:sdr}
    }
    \caption{Experimental setup. (a) GNSS receiving antenna and (b) SDR-based attack devices and victim receivers.}
    \label{fig:setup}
\end{figure}

We implemented the proposed attacks on OSNMA-enabled receivers in real-world environments, using a PC (Lenovo Y9000P), USRP B210 SDRs, and GNSS antennas.
The victim receivers include two commercial receivers: a Septentrio Mosaic X5 receiver and a U-blox ZED-F9P receiver, and two open-source SDR-based receivers: GNSS-SDR \cite{gnsssdr38} and OSNMAlib \cite{galan2025improving}, as shown in Fig. \ref{fig:setup}.
We evaluate whether each attack passes OSNMA authentication and whether it induces the intended position and time solution.

\subsection{TSR Attack Evaluation}
The TSR experiments investigates whether LRT manipulation makes outdated but authentic signals satisfy the TS check and produce replayed position and time solutions.
To implement the TSR attack, we used a PC and a USRP B210 device equipped with two antennas, one for recording authentic Galileo signals and the other for retransmitting them.  
Since both the GST and the LRT are manipulated, we used Coordinated Universal Time (UTC) provided by the National Institute of Standards and Technology (NIST) as an independent trusted time reference.

\begin{figure}[h]
\centerline{\includegraphics[width=\linewidth]{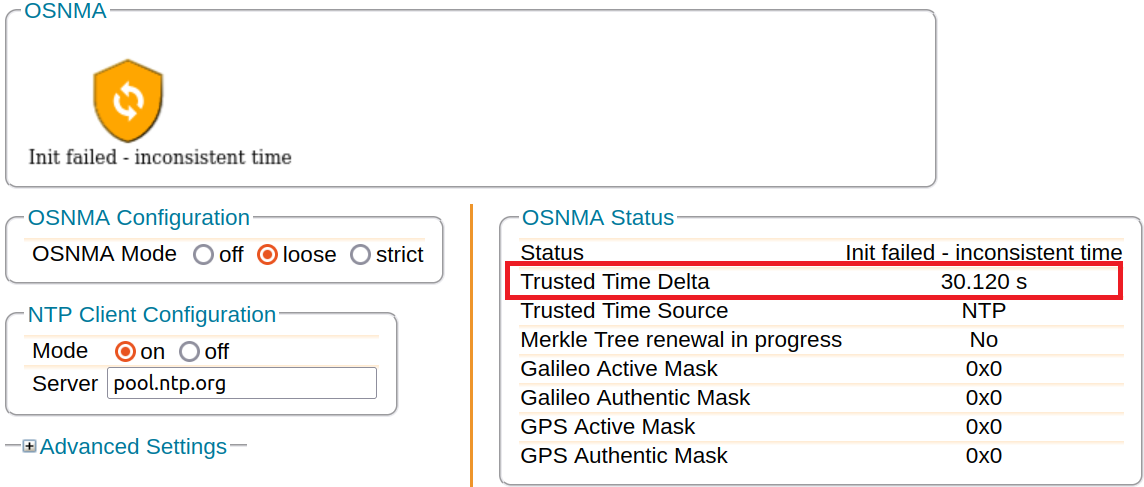}}
\caption{A failed replay attack on the Septentrio receiver without LRT manipulation.}
\label{fig:delay30}
\end{figure}
The Septentrio receiver obtains its LRT from public NTP servers.
To meet the TS requirements, we used the open-source Delorean tool \cite{delorean-3} to manipulate the receiver's LRT, aligning it with GST in the recorded signal.
The error bound $B$ in the TS rule is set to 30~s. 
As a baseline, when the replay delay exceeded 30~s without LRT manipulation, the difference between the replayed GST and the receiver's LRT exceeded $B$. 
Consequently, the receiver failed the TS check and did not initiate OSNMA verification, as shown in Fig. \ref{fig:delay30}.
\begin{figure}[h]
\centerline{\includegraphics[width=\linewidth]{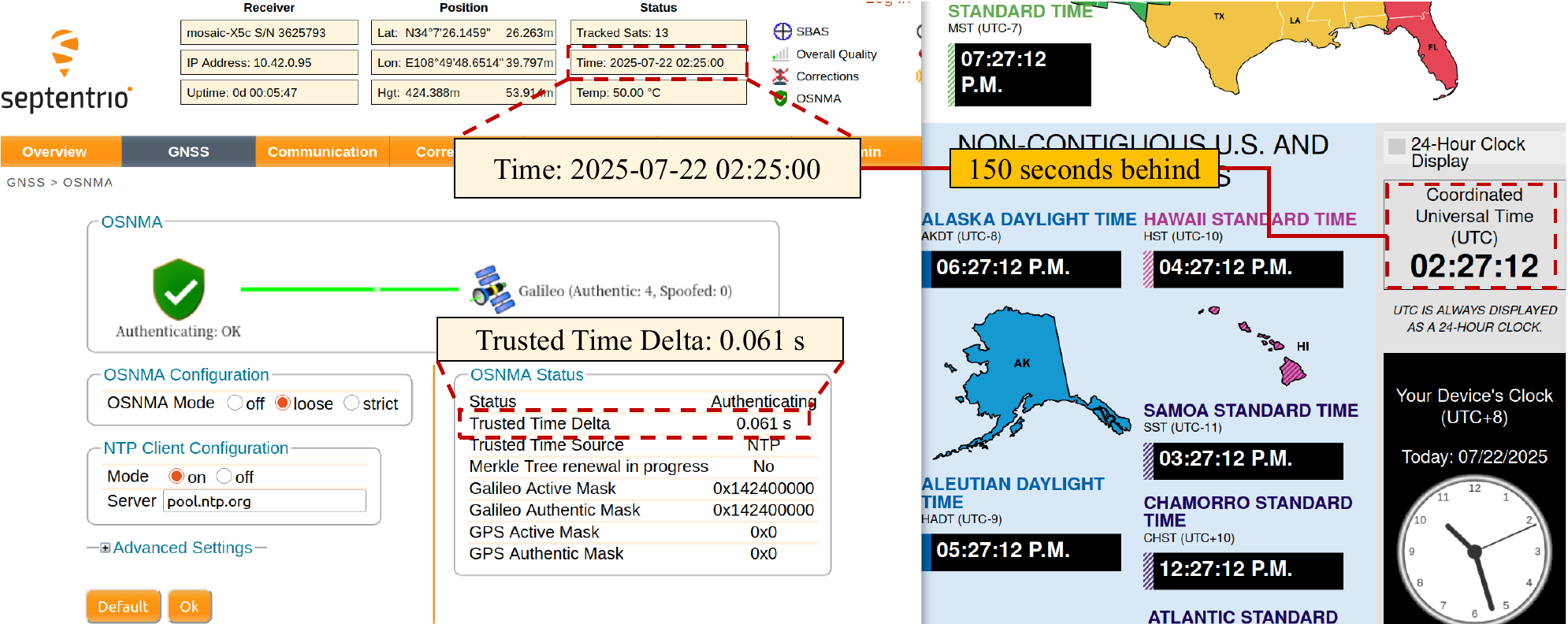}}
\caption{A successful TSR attack with a 150-second GST offset on the Septentrio receiver.}
\label{fig:delay150}
\end{figure}
We then manipulated the receiver's LRT to remain consistent with the outdated GST in the replayed signal.
Fig. \ref{fig:delay150} shows a successful TSR attack with a replay delay of approximately $150$ seconds.
The OSNMA view shows the current authentication status, the time difference between the GST and LRT (i.e., Trusted Time Delta) and the source used to obtain the LRT (i.e., Trusted Time Source) and NTP server address (pool.ntp.org).
The measured Trusted Time Delta was only 0.061~s, which is substantially smaller than $B=30~s$. 
The receiver therefore accepted the manipulated LRT and replayed GST as synchronized and completed OSNMA authentication.
The receiver's time solution was 2025-07-22 02:25:00, whereas the trusted UTC time was 2025-07-22 02:27:12.
At the time of the experiment, GST was $18$ seconds ahead of UTC.
This result implies that the time solution derived by the receiver lagged 150~s behind the trusted time.

\begin{figure}[h]
\centerline{\includegraphics[width=\linewidth]{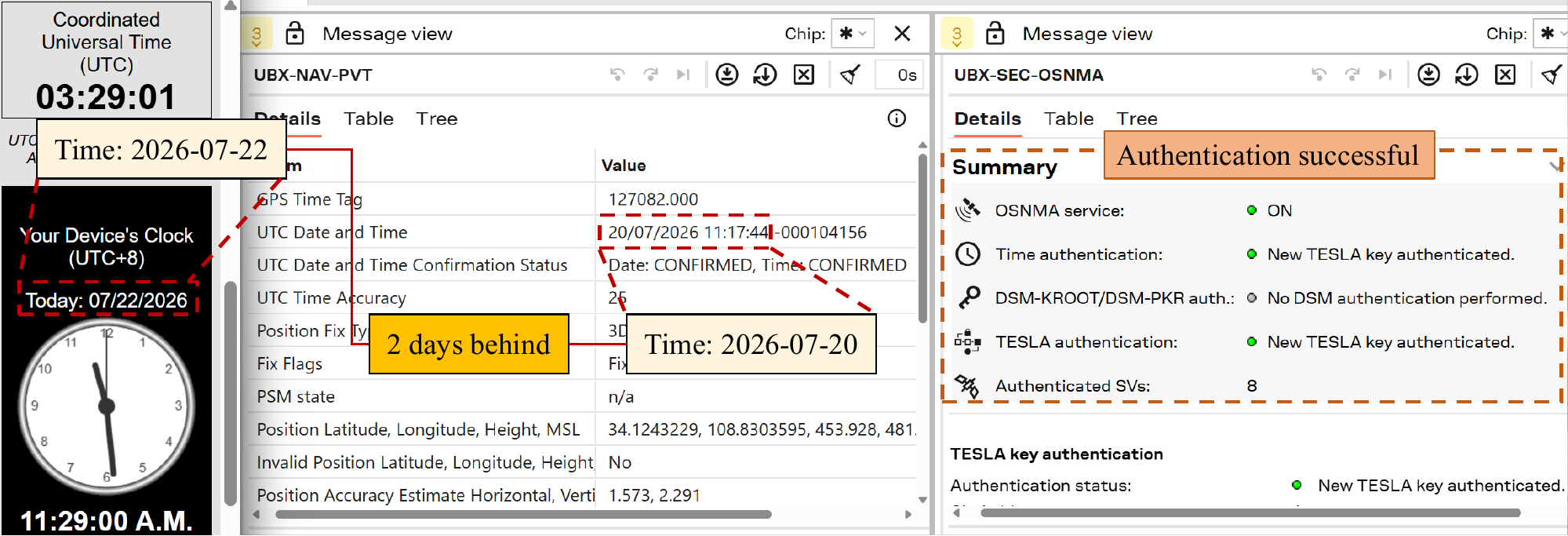}}
\caption{A successful TSR attack with a delay of more than two days on the U-blox receiver.}
\label{fig:delay2days}
\end{figure}
The U-blox receiver obtains its LRT either from the host PC's system time or via the UBX-MGA-INI-TIME-UTC command.
The default value of $B$ is 15 seconds. 
We manipulate the receiver's LRT by attacking the host time or by sending forged time commands to the receiver to directly tamper with the LRT.
Both methods aligned the LRT with the outdated GST.
Fig. \ref{fig:delay2days} shows a successful TSR attack with a replay delay of more than two days.
The UBX-SEC-OSNMA message reports successful OSNMA initialization, the root key, TESLA key, and MAC were authenticated. 
Comparison with the trusted NIST UTC shows that the receiver’s UTC solution lagged by more than two days.
The above results demonstrate that by manipulating the GST and LRT to satisfy TS requirements, TSR attack can pass OSNMA authentication and spoof the receivers to the position and past time contained in the recorded signals.

\begin{figure*}[t]
    \centering
    \subfigure[]{
        \includegraphics[width=0.48\linewidth]{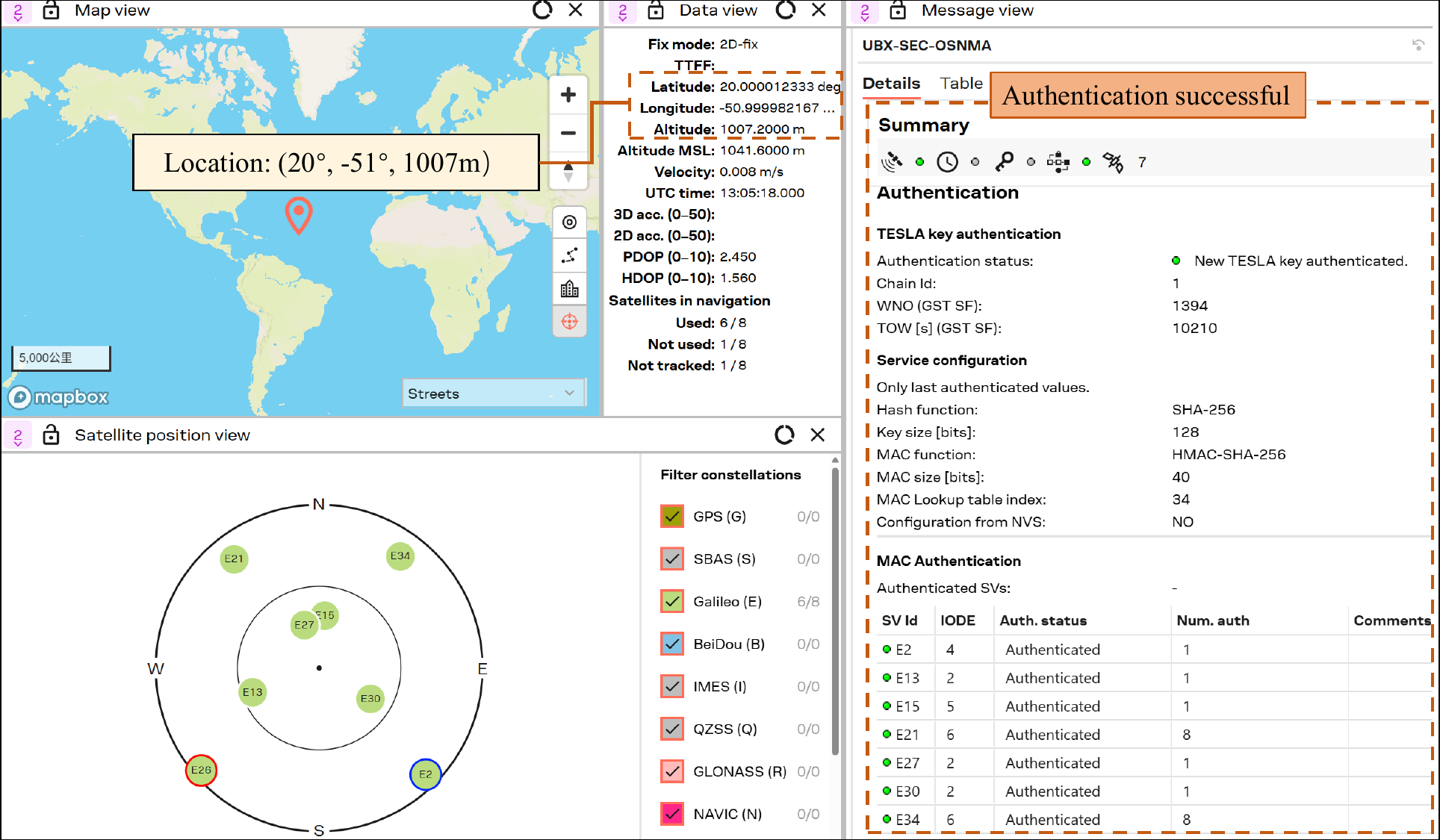}
        \label{fig:a}
    }
    \subfigure[]{
        \includegraphics[width=0.48\linewidth]{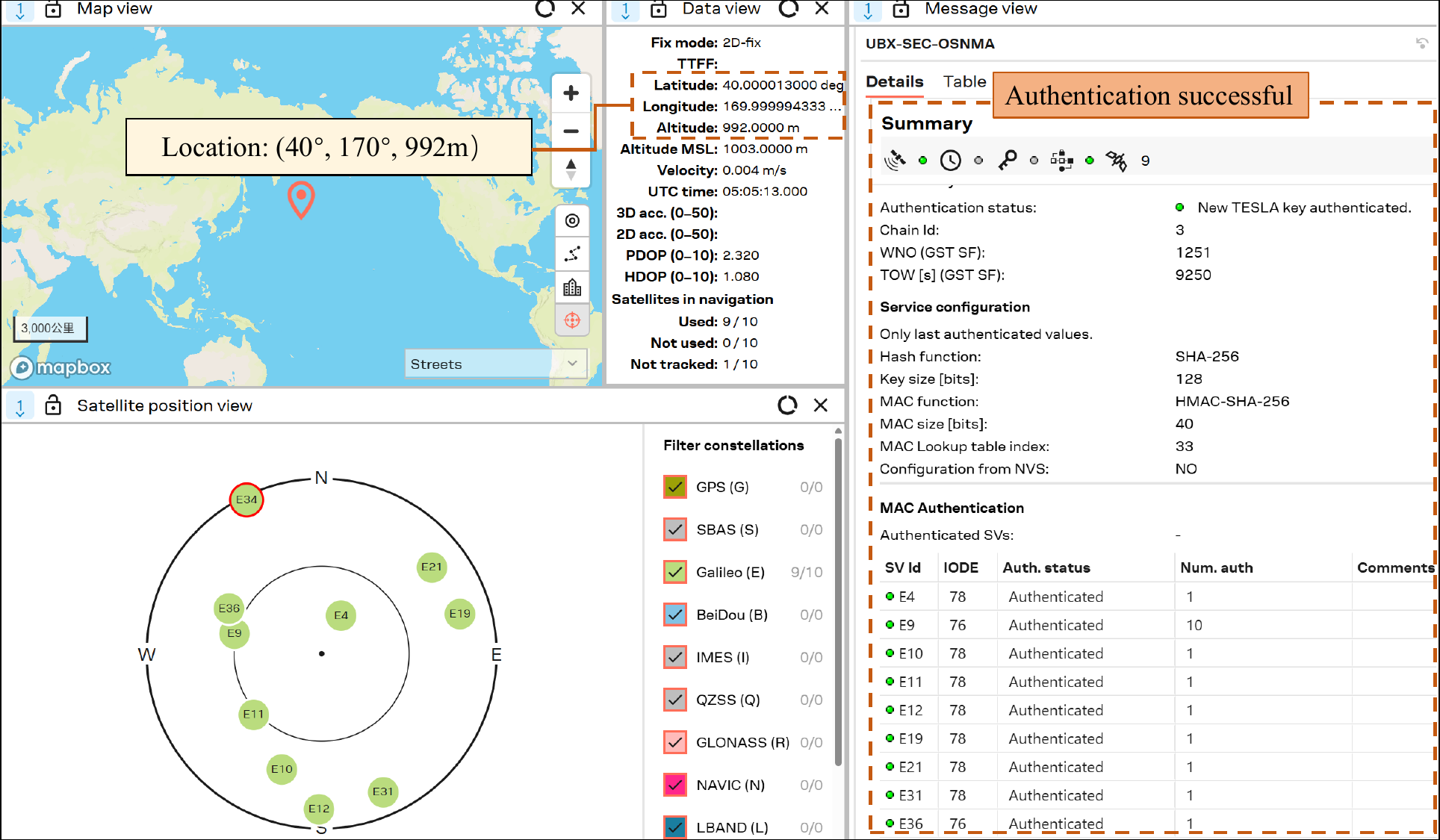}
        \label{fig:b}
    }
    \subfigure[]{
        \includegraphics[width=0.48\linewidth]{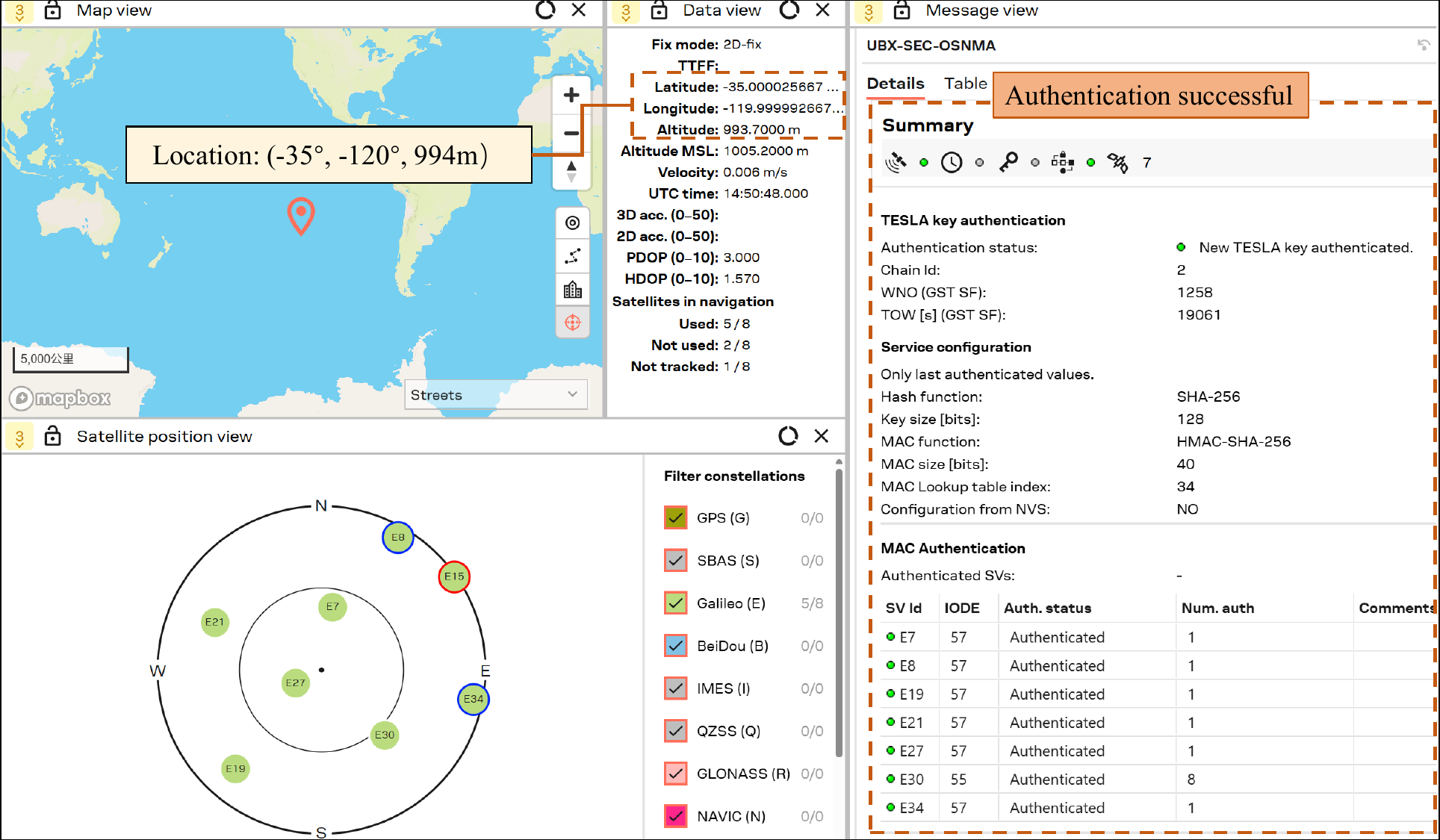}
        \label{fig:c}
    }
        \subfigure[]{
        \includegraphics[width=0.48\linewidth]{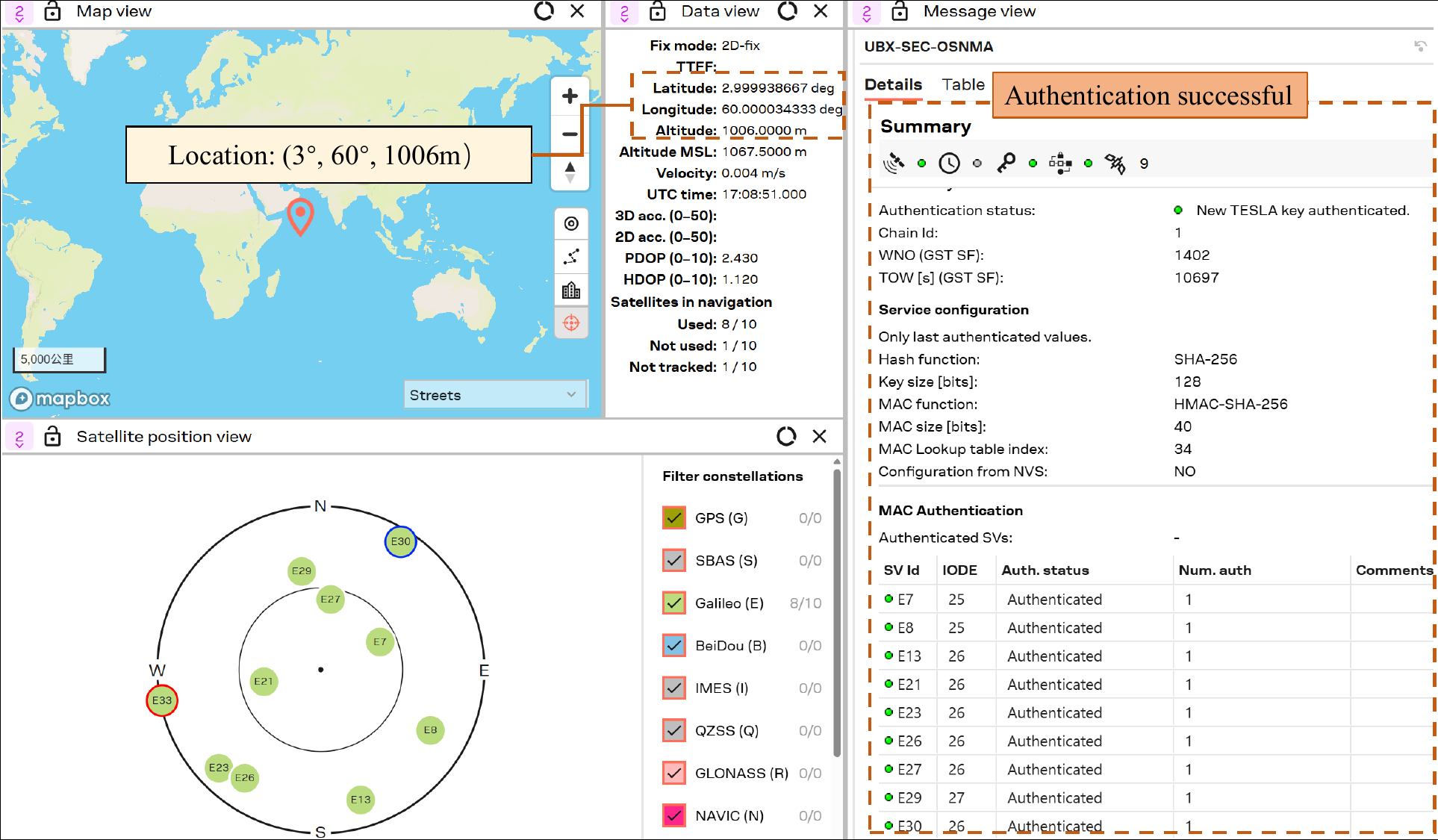}
        \label{fig:d}
    }
    \caption{Successful TSF attacks on the U-blox receiver at four target locations and times.}
    \label{fig:four}
\end{figure*}

\subsection{TSF Attack Evaluation}\label{sec:TSF impl}

To implement the TSF attack, we used one USRP B210 to transmit the forged signals.
In addition, we utilized a GNSS antenna mounted on the laboratory roof to extract auxiliary OSNMA data for different time intervals, as shown in Fig. \ref{fig:antenna}.
As a representative example, we configured the target position as ($20^\circ$, $-51^\circ$, $1000$ m), with a signal start time of 2026-05-13 13:00:00 and a duration of one hour.
Based on the steps introduced in Section~\ref{sec:TSF attack}, we conducted the TSF attack as follows.
\begin{enumerate}
	\item We determined the set of satellites visible from the target location as PRNs [2, 13, 15, 21, 27, 30, 34], and generated the navigation data, PRN code, code phase, and Doppler frequency for each satellite. 
    \item According to the configured signal start time and duration, we retrieved auxiliary OSNMA data covering the corresponding time interval. 
    From these data, we extracted the required OSNMA parameters, including the local TESLA key chain, MAC function (SHA-256) and tag size (40 bits). Furthermore, the extracted MACLT=34, representing the tag sequence is (00S, FLX, 04S, FLX, 12S, 00E).
    \item We selected PRNs [2, 15, 27] as satellites that did not transmit OSNMA data. 
    We then calculated their inter-satellite distances relative to OSNMA-transmitting PRNs [13, 21, 30, 34] for subsequent tag allocation. 
    \item Based on Eq. \ref{eq1}, we generated the tags for every subframe of each satellite using the forged navigation data and the local TESLA keys. 
    The generated tags included both self-authentication and cross-satellite authentication tags associated with different ADKD types. 
    \item We initialized the OSNMA fields for OSNMA-transmitting satellites and embedded the HKROOT messages and TESLA keys from the auxiliary OSNMA data. 
    Following the tag sequence, each self-authentication tag was embedded into the tag slot associated with the corresponding "S" element. 
    The cross-satellite tags were then embedded to the slots associated with the "E" and "FLX" elements in ascending order of inter-satellite distance. 
    The corresponding ADKD values and satellite PRNs were also inserted into the MACK message.
	\item We combined the constructed OSNMA fields with the forged navigation data, recomputed each page’s CRC, and assembled complete I/NAV messages.
    Finally, we applied Composite Binary Offset Carrier (CBOC) modulation using the navigation data, spreading codes, code phases, and Doppler frequencies. 
    The resulting baseband I/Q samples were transmitted by the USRP B210 at the Galileo E1 center frequency of 1.57542~GHz.
\end{enumerate}

Fig.~\ref{fig:a} presents the experimental results of a TSF attack against the U-blox receiver. 
As shown in the OSNMA status view, root key verification, TESLA key verification, and tag verification were all successfully completed. 
The receiver parsed the MACLT, MAC function, and MAC length, which are consistent with our settings.
Of the seven authenticated satellites, PRNs [2, 15, 27] do not transmit OSNMA data, but their associated tags are authenticated by the OSNMA-transmitting satellites through the cross-satellite authentication mechanism. 
This indicates that the receiver accepted the constructed OSNMA authentication relationships.
The satellite position view shows that all seven forged satellites were tracked by the receiver.
Using the authenticated navigation data, the receiver calculated a position solution of $(20^\circ, -51^\circ, 1007~\mathrm{m})$, which was consistent with the predefined target location. 
The map view further visualizes the receiver position after successful spoofing. 
Moreover, the data view shows that the receiver's UTC solution was shifted to the predefined target time.

To evaluate different satellite geometries and authentication epochs, we conducted repeated TSF attack experiments at three other target locations and start times.
These predefined locations and times were $(40^\circ, 170^\circ, 1000~\mathrm{m})$ at 2023-08-16 05:00:00, $(-35^\circ, -120^\circ, 1000~\mathrm{m})$ at 2023-10-17 14:45:00 and $(3^\circ, 60^\circ, 1000~\mathrm{m})$ at 2026-07-08 17:00:00.
Figs.~\ref{fig:b}--\ref{fig:d} show the receiver was successfully spoofed to the predefined positions $(40^\circ, 170^\circ, 992~\mathrm{m})$, $(-35^\circ, -120^\circ, 994~\mathrm{m})$, and $(3^\circ, 60^\circ, 1006~\mathrm{m})$. 
In each experiment, both the position and time solutions were consistent with their target values.
Fig. \ref{fig:four} further illustrates the changes in OSNMA configuration over time. For example, the TESLA key chain ID changes from 1 to 3, and MACLT changes from 33 to 34. 
The results show that the TSF attack remains effective despite these changes, and the forged signal consistently passes the receiver's OSNMA authentication.

\subsection{TSDF Attack Evaluation}
\begin{figure}[t]
\centerline{\includegraphics[width=\linewidth]{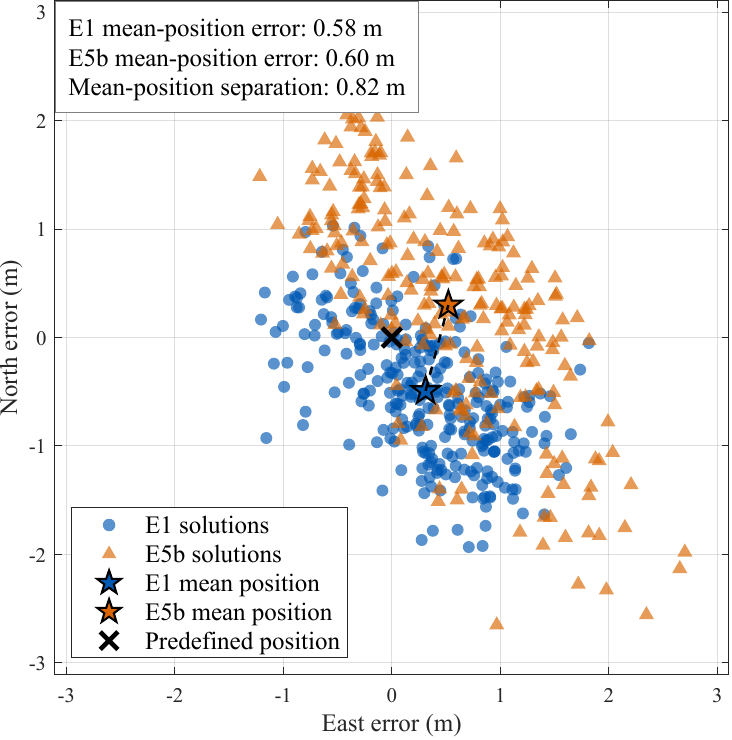}}
\caption{Horizontal positioning error distributions for the independently
forged E1 and E5b signals.}
\label{fig:scatter}
\end{figure}
A single USRP B210 cannot transmit E1 and E5b signals simultaneously because the device's bandwidth cannot cover both signals.
Therefore, we used two USRP B210s to transmit the forged E1 and E5b signals and two additional B210s to receive the respective bands.
The PC ran SDR-based software receivers, GNSS-SDR and OSNMAlib, both of which use NTP to obtain LRT.
GNSS-SDR computed the PVT solutions and forwarded the extracted E1 and E5b navigation messages to OSNMAlib for cross-band authentication.
The predefined target location, start time, and duration are the same as in Section \ref{sec:TSF impl}.

\begin{figure}[h]
\centerline{\includegraphics[width=\linewidth]{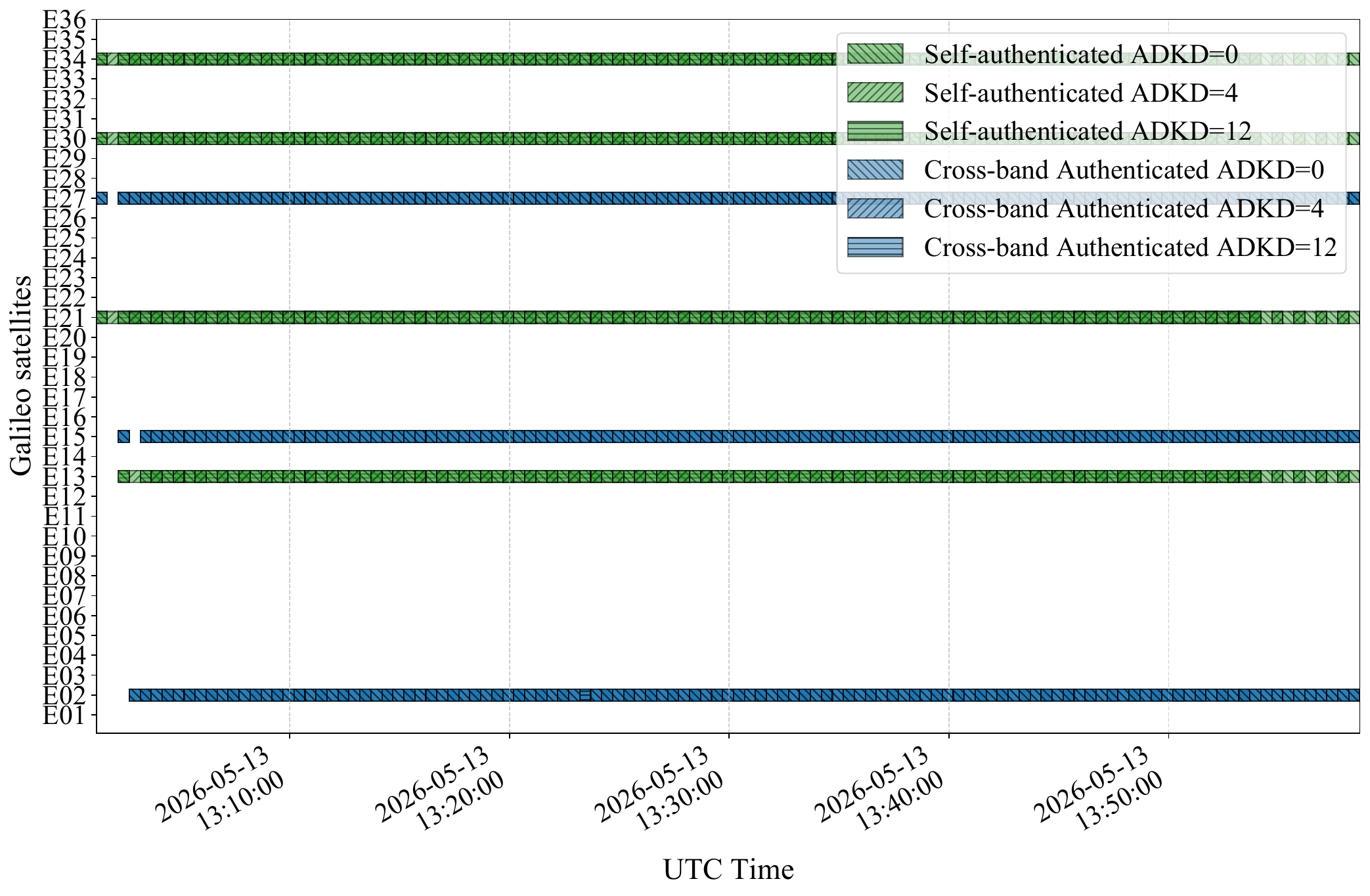}}
\caption{OSNMA authentication timeline during the one-hour TSDF attack.}
\label{fig:authen-timeline}
\end{figure}

Fig. \ref{fig:scatter} shows position solutions obtained independently from the forged E1 and E5b signals, where the target position is set to ($20^\circ$, $-51^\circ$, $1000$ m).
The mean horizontal position errors of E1 and E5b are 0.58~m and 0.60~m, respectively. Moreover, the horizontal separation between the two mean horizontal positions is only 0.82~m. 
The close position solutions demonstrate cross-band measurement consistency.
Furthermore, OSNMAlib successfully authenticated the E1 tags associated with the E5b navigation data. 
We utilized the OSNMA authentication timeline to illustrate the authentication status of both self-authentication tags and cross-band tags, as shown in Fig. \ref{fig:authen-timeline}. 
During the one-hour TSDF attack, at least four satellites were consistently authenticated, providing support for continuous authenticated positioning.

\subsection{Practical Feasibility under Transmit Gain and Distance}
We investigated the boundary conditions for successful TSR and TSF attacks by adjusting the USRP transmit gain and the distance between the attacker and the victim's receiver.
The feasibility results of the TSF attack are summarized in Table~\ref{table-TSF}, with 10 experiments repeated for each gain and distance.
When the gain is set to 70~dB, successful spoofing can be achieved within a distance of up to 4~m.
In contrast, when the gain is below 20~dB, the TSF attack consistently fails. 

\begin{table}[h]
	\centering
	\caption{Feasibility of TSF attacks under different distances and transmission power. '-' denotes inconsistent results from repeated experiments.}
	\label{table-TSF}
	\begin{tabular}{|c|c|c|c|c|c|c|c|c|c|c|}
		\hline
		\diagbox[width=6em]{\makecell[cc]{Gain\\(dB)}}{\makecell[cc]{Distance\\(m)}}& 1  & 2 & 3 & 4 & 5 & 6 & 7 & 8 & 9 & 10  \\
		\hline
		70& \checkmark & \checkmark  & \checkmark & \checkmark & -  & - & $\times$ & $\times$ & $\times$ & $\times$
		\\
		\hline
		60&\checkmark & \checkmark  & - & $\times$ & $\times$  & $\times$ & $\times$ & $\times$ & $\times$ & $\times$\\
		\hline
		50& \checkmark & -  & $\times$ & $\times$ & $\times$  & $\times$ & $\times$ & $\times$ & $\times$ & $\times$
		\\
		\hline
		40&- & -  & $\times$ & $\times$ & $\times$  & $\times$ & $\times$ & $\times$ & $\times$ & $\times$
		\\
		\hline
		30&- & $\times$  & $\times$ & $\times$ & $\times$  & $\times$ & $\times$ & $\times$ & $\times$ & $\times$
		\\
		\hline
		20&$\times$ & $\times$  & $\times$ & $\times$ & $\times$  & $\times$ & $\times$ & $\times$ & $\times$ & $\times$
		\\
		\hline
		10&$\times$ & $\times$  & $\times$ & $\times$ & $\times$  & $\times$ & $\times$ & $\times$ & $\times$ & $\times$  \\
		\hline
	\end{tabular}
\end{table}

The results showed that as gain increases or attack distance decreases, the feasibility of the TSF attack increases, and TSR showed the same trend.
Note that these feasibility boundaries are closely related to hardware configuration. 
For example, equipping the attacker with higher-gain antennas or more powerful transmission hardware could further extend the effective attack range.

\section{Discussion} \label{sec:discuss}
\subsection{OSNMA Enhancements}
\subsubsection{Defense Against ATS-based Attacks}
The receiver should use trusted time source, reject implausible backward jumps, and disable OSNMA authentication if the time source is untrusted.
Enhancement measures depend on how receivers obtain the LRT.
\begin{itemize}
	\item \emph{Internal RTC}: A smaller $B$ provides better spoofing mitigation but reduces the usability of OSNMA because the LRT needs to be frequently calibrated. 
    Finding the optimal $B$ is thus a trade-off problem between security and usability.
    Given an RTC with daily timing error $t_{error}$, the number of days a receiver can safely perform OSNMA without recalibration is $N_{safe} \approx {B}/{t_{error}}$. Thus, given a constraint  $N_{safe}\geq N$, the optimal $B$ is $N t_{error}$.
    In practice, using high-stability oscillators such as TCXO or OCXO \cite{OSNMA-oscillators} can help achieve less frequent calibration, thus allowing manufacturers to set smaller $B$.
    \item \emph{Network time}:
    Receivers synchronized via a network timing server typically have timing error within a few hundred milliseconds, and this error does not accumulate over time.
    Since the traditional NTP is vulnerable to time spoofing, receiver manufacturers should adopt more secure Network Time Security (NTS) protocol \cite{Ntp-security45} and update existing implementations where possible.
    \item \emph{Host or receiver commands}: The time setting interface should require authenticated and authorized access. 
    Receivers should compare the acquired LRT with other available time sources to avoid large or backward corrections. 
\end{itemize}
\subsubsection{Geography-Bound Keys}
In the current OSNMA design, all satellites broadcast the same key at the same GST.
Once an attacker obtains a disclosed key from a satellite, this key can be reused to forge OSNMA data for all satellites and spoof a receiver to any location on Earth. 
If keys are bound to geography, i.e., the satellites covering a certain area broadcast the same key at the same GST, as mentioned in \cite{osnma8-2016}, TSF attacks would be restricted to only this area.
\subsection{Ethical Disclosure}
To ensure safety, all spoofing experiments were performed in a controlled environment designed to prevent interference with operational GNSS users and systems.
We responsibly disclosed the identified ATS and our attack experiments to the relevant parties through the Septentrio Support Portal \cite{septentrio-problem}, the U-blox Support Community \cite{ubloxportal}, and the GSC Help Desk \cite{GSC}.

\section{Conclusion}\label{sec:conclusion}
We identified and experimentally demonstrated an exploitable artificially manipulated time synchronization (ATS) condition in OSNMA-enabled receivers.
Under this condition, attackers can jointly manipulate the receiver's LRT and the GST in the spoofing signal, forcing the receiver to satisfy the Time Synchronization (TS) requirements.
Exploiting the ATS condition, we proposed a TS-compliant spoofing framework against OSNMA-protected receivers, including TS-compliant replay, TS-compliant forgery, and TS-compliant dual-frequency forgery attacks. 
We conducted extensive practical experiments on two commercial receivers and two SDR-based receivers to verify the feasibility of these attacks.
The results showed that these attacks can pass OSNMA authentication and spoof the receiver to arbitrary locations and past time for which the required authentication material is available.
We further discussed OSNMA enhancement strategies at the receiver and system level to defend against the proposed attacks.


\bibliographystyle{ieeetr}
\bibliography{ref}

\end{document}